\documentclass[aps,prb,amsmath,amssymb,twocolumn]{revtex4}
\usepackage{graphicx}
\usepackage{bm}
\usepackage{bbold}

\def\bk{ \bm{k} }

\def\bgam{ \bm{\gamma} }

\def\sgn{\, \mathrm{sgn}\, }

\begin{document}
\title{Stability of the boundary zero modes in one-dimensional topological superconductors}

\author{K. V. Samokhin and B. P. Truong}

\affiliation{Department of Physics, Brock University, St. Catharines, Ontario L2S 3A1, Canada}
\date{\today}

\begin{abstract}
We calculate the spectrum of the Andreev bound states in a one-dimensional superconductor with a strong Rashba spin-orbit coupling. We focus on the fate 
of the zero-energy Andreev modes in the presence of time reversal symmetry-breaking perturbations, both at the boundary and in the bulk. It is shown that the zero modes are destroyed by
time reversal symmetry-breaking fluctuations, even if the mean-field state of the system is time-reversal invariant and topologically nontrivial.
\end{abstract}

\maketitle

\section{Introduction}
\label{sec: Intro}

For more than three decades, topological quantum systems have remained at the centre of attention in condensed matter physics.\cite{top-SC} These diverse systems, from a two-dimensional (2D) electron gas 
exhibiting the integer Quantum Hall Effect\cite{TKNN82}, to chiral $p$-wave superconductors and superfluids,\cite{Volovik-book} to topological band insulators,\cite{KM05,TI-review} all share one common feature: 
the quantum states in the bulk fall into distinct classes characterized by various topological invariants, which are robust under sufficiently small perturbations. The choice of the bulk invariant is determined by 
the symmetry and dimensionality of the system, with a particularly important role played by time reversal symmetry (TRS). 
According to Ref. \onlinecite{SRFL08}, for single-particle Hamiltonians the topological invariant either takes an
integer value or is a $\mathbb{Z}_2$ quantity, although this may be significantly modified by interactions.\cite{interactions-Z8}   

There is a widely held belief that a nontrivial topology of the bulk manifests itself in the presence of protected gapless quasiparticle states localized near the boundary of the system, which is known as 
the bulk-boundary correspondence.\cite{Volovik-book} Archetypal examples include the current-carrying chiral edge states in the quantum Hall\cite{QHE-edge} 
or topological band insulators,\cite{KM05} and also the Andreev bound states (ABS) in the chiral $p$-wave\cite{ABS-pwave} and the nodal $d$-wave superconductors.\cite{Hu94} 
In some cases, it is possible to obtain an explicit analytical relation between the number of the zero-energy modes and a certain topological invariant in the bulk.\cite{BBC} For instance, the 
quantum-Hall edge states are related to the first Chern number of the 2D Brillouin zone, which is also known as the TKNN integer, after Ref. \onlinecite{TKNN82}, 
while the dispersionless surface ABS in TR invariant unconventional
superconductors are controlled by the phase winding number of the determinant of the off-diagonal Bogoliubov-de Gennes (BdG) Hamiltonian.\cite{STYY11} Note, however, that there are some recent results\cite{MR15} 
that call into question the universality of the bulk-boundary correspondence, at least in the $\mathbb{Z}_2$ case. 

The extent to which the zero modes are ``protected'', i.e., insensitive to the boundary details, is one of the less-understood aspects of the bulk-boundary correspondence. Given the crucial role played by TRS, 
of particular interest here is the fate of the zero modes in the situations when the bulk is TR invariant, while the boundary is not. The goal of this paper is to study the effects on the ABS of (i) the magnetic boundary 
scattering and (ii) an intrinsic TRS breaking in the superconducting state, both at the mean-field level and also including fluctuations. 
One can capture the essential physics by looking at a half-infinite one-dimensional (1D) superconductor in contact with a ferromagnetic insulator. In contrast to previous works, see Ref. \onlinecite{TRB-1D}, we do not 
consider the TRS breaking by an applied magnetic field. 
 
The 1D (or rather, quasi-1D) superconductivity can be realized in a metallic quantum wire on a substrate. The pairing interaction can be either intrinsic 
to the wire or extrinsic, i.e., induced by the substrate. Inversion symmetry is absent in this system and a crucial role is played by the Rashba spin-orbit (SO) coupling of the electrons in the wire 
with the asymmetric substrate potential, see Refs. \onlinecite{Rashba-model}, \onlinecite{Manchon15}, and the references therein. The Rashba SO coupling lifts the spin degeneracy of the electron states, 
producing nondegenerate Bloch bands labeled by ``helicity'', with the wave functions characterized by a nontrivial momentum-space topology. 
Its profound consequences for superconductivity have been extensively studied in the last decade, see Refs. \onlinecite{NCSC-book}, \onlinecite{Kneid15}, and \onlinecite{Smid17} for reviews.    

The paper is organized as follows. The structure of the single-electron bands and the superconducting pairing in a 1D Rashba wire are discussed in Sec. \ref{sec: electron bands}. In Sec. \ref{sec: ABS}, we present
the semiclassical derivation of the ABS spectrum, which is expressed in terms of the boundary scattering matrix. The latter is calculated in Sec. \ref{sec: FM boundary}. Stability of the ABS zero modes against TRS-breaking 
perturbations is studied in Sec. \ref{sec: ABS-stability}.
Throughout the paper we use the units in which $\hbar=k_B=1$, neglecting, in particular, the difference between the quasiparticle momentum and wave vector.

\section{Superconductivity in 1D nondegenerate bands}
\label{sec: electron bands}

We consider a quasi-1D electron gas on a $xy$-plane substrate. Neglecting the lattice periodicity, the three-dimensional (3D) potential $U(x,y,z)$ affecting the electrons is constant in $x$ direction, 
but confining in both $y$ and $z$ directions. This system is TR invariant in the normal state but lacks an inversion center, because the substrate breaks the $z\to-z$ mirror reflection symmetry. 
The momentum space is one-dimensional, labelled by the wave vector $\bk=k_x\hat{\bm{x}}$, where $-\infty<k_x<\infty$. 

The simplest Hamiltonian that captures the essential features of the electronic band structure in a noncentrosymmetric 1D system has the following form:
\begin{equation}
\label{H-Rashba}
    \hat H_0=\sum\limits_{k_x}\sum_{s,s'=\uparrow,\downarrow}\left[\epsilon_0(k_x)\delta_{ss'}+\bgam(k_x)\bm{\sigma}_{ss'}\right]\hat b^\dagger_{k_x,s}\hat b_{k_x,s'}.
\end{equation}
This is the 1D version of the well-known Rashba model.\cite{Rashba-model} The first term describes a single spin-degenerate band, for which we use the effective mass approximation:
\begin{equation}
\label{eff-mass}
  \epsilon_0(k_x)=\frac{k_x^2}{2m^*}-\epsilon_F,
\end{equation}
where $\epsilon_F=k_F^2/2m^*$ is the Fermi energy (the difference between $\epsilon_F$ and the chemical potential is neglected). The second term in Eq. (\ref{H-Rashba}) is the asymmetric SO coupling, with 
$\hat{\bm{\sigma}}$ being the Pauli matrices.
While the momentum space is 1D, the spin space is still 3D, so that the asymmetric SO coupling is described by the 3D pseudovector $\bgam(k_x)$, 
which is real and odd in $k_x$ due to the TRS. The simplest expression compatible with both requirements is 
\begin{equation}
\label{gamma-Rashba}
  \bgam(k_x)=\bm{a}k_x,
\end{equation}
with a real $\bm{a}$. In the absence of additional mirror reflection symmetries of the confining potential,\cite{Sam17}  there are no further constraints on the components of $\bm{a}$.
Below we will use the spherical angle parameterization,
\begin{equation}
\label{a-angles}
  \bm{a}=|\bm{a}|(\sin\alpha\cos\beta,\sin\alpha\sin\beta,\cos\alpha),
\end{equation} 
where $0\leq\alpha\leq\pi$ and $0\leq\beta<2\pi$.

\begin{figure}
\includegraphics[width=7cm]{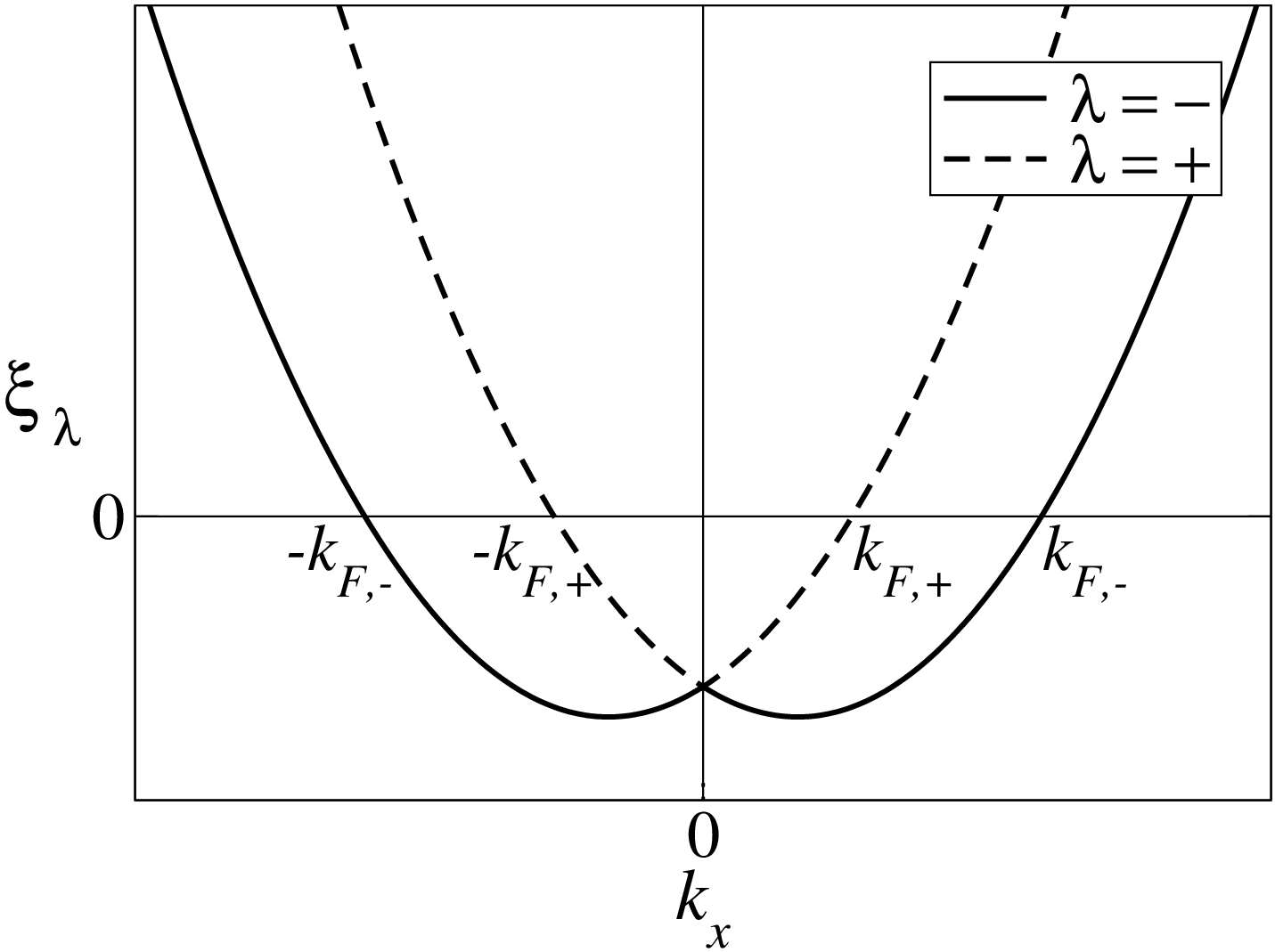}
\caption{The 1D helicity bands and the Fermi points.}
\label{fig: bands}
\end{figure}

The asymmetric SO coupling lifts the spin degeneracy of the bands almost everywhere in the momentum space. Diagonalizing Eq. (\ref{H-Rashba}), we obtain:
\begin{equation}
\label{Rashba-bands}
    \xi_\lambda(k_x)=\frac{k_x^2-k_F^2}{2m^*}+\lambda|\bm{a}||k_x|=\xi_\lambda(-k_x).
\end{equation}
Here the band index $\lambda=\pm$, called the helicity, has the meaning of the spin projection on the direction of motion. Although the two 1D Fermi ``surfaces'', defined by the equations $\xi_\pm(k_x)=0$,
see Fig. \ref{fig: bands}, have different sizes: 
\begin{equation}
\label{Fermi-wavevectors}
  k_{F,\lambda}=\tilde k_F-\lambda\Lambda,\quad \tilde k_F=\sqrt{k_F^2+\Lambda^2}, 
\end{equation}
where $\Lambda=m^*|\bm{a}|$, the Fermi velocities are the same in both bands and equal to $\tilde v_F=\tilde k_F/m^*$.
The corresponding eigenstates can be chosen in the following form:
\begin{eqnarray}
\label{Rashba-eigenstates}
  |k_x,+\rangle=\frac{1}{\sqrt{2\tilde v_F}}\left(\begin{array}{c}
                                        e^{-i\beta/2}\sqrt{1+\sgn k_x\cos\alpha}\vspace*{5pt} \\
					e^{i\beta/2}\sgn k_x\sqrt{1-\sgn k_x\cos\alpha}
                                        \end{array}\right),\nonumber\\ \\
   |k_x,-\rangle=\frac{1}{\sqrt{2\tilde v_F}}\left(\begin{array}{c}
                                        -ie^{-i\beta/2}\sqrt{1-\sgn k_x\cos\alpha}\vspace*{5pt} \\
					ie^{i\beta/2}\sgn k_x\sqrt{1+\sgn k_x\cos\alpha}
                                        \end{array}\right),\nonumber
\end{eqnarray}
normalized to produce the current of unit magnitude. 

The TR invariance of the normal state implies that the states $|k_x,\lambda\rangle$ and $K|k_x,\lambda\rangle$ have the same energy, forming a Kramers doublet 
(recall that the TR operator for spin-1/2 particles is $K=i\hat\sigma_2K_0$, where $\hat\sigma_2$ is the Pauli matrix and $K_0$ is complex conjugation). Therefore,
\begin{equation}
\label{t-factor}
  K|k_x,\lambda\rangle=t_\lambda(k_x)|-k_x,\lambda\rangle,
\end{equation}
where $t_\lambda$ is a phase factor.\cite{t-factor} Since $K^2=-1$ for fermions, one can show that $t_\lambda(-k_x)=-t_\lambda(k_x)$.
Note that the phase factors are not gauge invariant, and for the eigenstates given by Eq. (\ref{Rashba-eigenstates}) have a particularly simple form:
\begin{equation}
\label{t-sgn-x}
  t_\lambda(k_x)=\sgn k_x.
\end{equation}
The helicity bands (\ref{Rashba-bands}) are nondegenerate at all $k_x$, except the TR invariant point $k_x=0$, where the SO coupling (\ref{gamma-Rashba}) vanishes and neither the eigenstates
nor the TR phase factors are well defined. In a lattice model taking into account the momentum space periodicity, there are additional TR invariant points at the 1D Brillouin zone boundaries, where the SO coupling also vanishes.  

Next, we use the helicity basis to construct the superconducting Hamiltonian. The only assumption we make is that, whatever the microscopic pairing mechanism, the SO band splitting is large enough
to suppress the pairing of electrons from different bands. In the mean-field approximation we have 
\begin{eqnarray}
\label{H-MF}
     \hat H &=& \sum_{k_x,\lambda}\xi_\lambda(k_x)\hat c^\dagger_{k_x,\lambda}\hat c_{k_x,\lambda}\nonumber\\
     &&+\frac{1}{2}\sum_{k_x,\lambda}\left[\Delta_\lambda(k_x)\hat c^\dagger_{k_x,\lambda}\hat{\tilde c}^\dagger_{k_x,\lambda}+\mathrm{H.c.}\right],
\end{eqnarray}
where the second term represents the intraband Cooper pairing between the states $|k_x,\lambda\rangle$ and $K|k_x,\lambda\rangle$. According to Eq. (\ref{t-factor}), the electron creation operator in 
the time-reversed state $K|k_x,\lambda\rangle$ is given by
\begin{equation}
\label{c-tilde c}
  \hat{\tilde c}^\dagger_{k_x,\lambda}\equiv K\hat c^\dagger_{k_x,\lambda}K^{-1}=t_\lambda(k_x)\hat c^\dagger_{-k_x,\lambda}.
\end{equation}
Substituting this last expression in $\hat H$ and using the anticommutation of the fermionic creation and annihilation operators, we obtain:
\begin{equation}
\label{Delta-even}
  \Delta_\lambda(k_x)=\Delta_\lambda(-k_x),
\end{equation}
therefore the pairing in the helicity representation is necessarily even in momentum. The TR operation acting on the gap functions is equivalent to the complex conjugation, i.e., $\Delta_\lambda(k_x)\to\Delta^*_\lambda(k_x)$,
while under an arbitrary rotation of the band state phases, $|k_x,\lambda\rangle\to e^{i\theta_\lambda(k_x)}|k_x,\lambda\rangle$, the gap functions remain invariant. 

In a BCS-type model, the gap functions are nonzero only in the vicinity of the Fermi level, where their $k_x$-dependence can be neglected. In our case, the 1D ``Fermi surface'' is given by the four Fermi wave vectors
$\pm k_{F,\pm}$, see Fig. \ref{fig: bands}, and the superconductivity is described by two complex order parameters $\Delta_+$ and $\Delta_-$. 
The stable uniform states are found by minimizing the Ginzburg-Landau free energy. 
In addition to the TR invariant states, in which the phases of the gap functions are either $0$ or $\pi$, there is also a phenomenological possibility of TRS-breaking states, both in the multiband and the two-band cases,
\cite{TRSB-states} in which
\begin{equation}
\label{gap-functions}
  \Delta_+=|\Delta_+|e^{i\chi},\quad\Delta_-=|\Delta_-|,
\end{equation}
where $0\leq\chi\leq\pi$ is the interband phase difference. The bulk quasiparticle spectrum consists of two electron-hole symmetric branches in each helicity band, given by $\pm\sqrt{\xi_\lambda^2+|\Delta_\lambda|^2}$, 
with the energy gap equal to $|\Delta_\lambda|$. 

Suppose the gap functions are real. Then, our system belongs to the symmetry class DIII in 1D and, therefore, can be characterized in the bulk by a $\mathbb{Z}_2$ 
topological invariant.\cite{SRFL08} Due to TR invariance, the boundary zero modes can only come in pairs,\cite{TRI-1D} and the $\mathbb{Z}_2$ invariant is nothing but the parity of the number of such Kramers pairs. 
For the Rashba superconductor with just two bands, this invariant has the following form:
\begin{equation}
\label{Z2-invariant}
  N_{1D}=\sgn\Delta_+\sgn\Delta_-,
\end{equation}
see Ref. \onlinecite{QHZ10}. The states with $N_{1D}=1$ ($\chi=0$) are topologically trivial, while those with $N_{1D}=-1$ ($\chi=\pi$) are nontrivial and should have two ABS zero modes. 
Below, in Secs. \ref{sec: ABS} and \ref{sec: ABS-stability}, we examine the validity of this last statement using an explicit calculation of the ABS energy under general assumptions about the boundary scattering.

\subsection{Spin representation}
\label{sec: Kitaev-chain}

It is instructive to translate the above results about the superconducting pairing in the helicity representation into the spin representation. The fermionic creation and annihilation operators are transformed into the spin basis 
using the following relations:
\begin{eqnarray*}
  \hat c^\dagger_{k_x,\lambda}=\sum_{s=\uparrow,\downarrow}\langle k_x,s|k_x,\lambda\rangle\hat b_{k_x,s}^\dagger,\\ 
  \hat{\tilde c}^\dagger_{k_x,\lambda}=\sum_{s=\uparrow,\downarrow}\langle k_x,\lambda|k_x,s\rangle\hat{\tilde b}_{k_x,s}^\dagger,
\end{eqnarray*}
where $\hat{\tilde b}_{k_x,s}^\dagger\equiv K\hat b_{k_x,s}^\dagger K^{-1}=(i\hat\sigma_2)_{ss'}\hat b_{-k_x,s'}^\dagger$. Substituting these expressions into
the pairing Hamiltonian (\ref{H-MF}), we obtain: 
\begin{eqnarray}
\label{H-MF-spin}
     \hat H &=& \sum_{k_x,ss'}\epsilon_{ss'}(k_x)\hat b_{k_x,s}^\dagger\hat b_{k_x,s'}\nonumber\\
     &&+\frac{1}{2}\sum_{k_x,ss'}\bigl[\Delta_{ss'}(k_x)\hat b_{k_x,s}^\dagger\hat b_{-k_x,s'}^\dagger+\mathrm{H.c.}\bigr].
\end{eqnarray}
The normal-state energy and the gap function become $2\times 2$ spin matrices: 
\begin{eqnarray}
\label{spin-matrices}
  && \hat\epsilon(k_x)=\sum_\lambda\xi_\lambda(k_x)\hat\Pi_\lambda(k_x),\nonumber\\ \\
  && \hat\Delta(k_x)=\sum_\lambda\Delta_\lambda(k_x)\hat\Pi_\lambda(k_x)(i\hat\sigma_2),\nonumber
\end{eqnarray}
where $\hat\Pi_\lambda(k_x)=|k_x,\lambda\rangle\langle k_x,\lambda|$ is the projection operator onto the $\lambda$th band.

From Eq. (\ref{Rashba-eigenstates}) we obtain the following expression for the projection operator:
$$
  \hat\Pi_\lambda(k_x)=\frac{1+\lambda\hat{\bgam}(k_x)\hat{\bm{\sigma}}}{2},
$$
where $\hat{\bgam}=\bgam/|\bgam|=\hat{\bm{a}}\sgn k_x$, see Eq. (\ref{gamma-Rashba}). Then, taking into account Eq. (\ref{Rashba-bands}), the normal-state energy matrix in Eq. (\ref{spin-matrices}) reproduces the Rashba 
Hamiltonian (\ref{H-Rashba}), while the gap function in the spin representation takes the form
\begin{equation}
\label{gap-spin}
  \hat\Delta(k_x)=\Delta_s(i\hat\sigma_2)+\Delta_t\hat{\bgam}(k_x)(i\hat{\bm{\sigma}}\hat\sigma_2).
\end{equation}
Here 
$$
  \Delta_s=\frac{\Delta_++\Delta_-}{2},\quad \Delta_t=\frac{\Delta_+-\Delta_-}{2},
$$
and we neglected the momentum dependence of $\Delta_\pm$. The expression (\ref{gap-spin}) describes a mixture of spin-singlet ($\Delta_s$) and spin-triplet ($\Delta_t$) pairing. The latter, which is characterized by the 
spin vector $\bm{d}(k_x)=\Delta_t\hat{\bgam}(k_x)$, is protected against the pair breaking effect of the Rashba SO band splitting.\cite{NCSC-book}

In a purely triplet superconducting state with $\Delta_+=-\Delta_-=\Delta$, we have $\hat\Delta(k_x)=\Delta\hat{\bgam}(k_x)(i\hat{\bm{\sigma}}\hat\sigma_2)$. Assuming for simplicity that 
$\bm{a}\parallel\hat{\bm{y}}$ in Eq. (\ref{gamma-Rashba}), which is required by symmetry for some 1D point groups,\cite{Sam17} we obtain: $\hat\Delta(k_x)=i\Delta\sgn k_x\hat\sigma_0$.
This gap function describes an odd in $k_x$ (``$p$-wave'') pairing state, which is TR invariant (because one can make the gaps real by a gauge transformation) and topologically nontrivial in the bulk, with two zero-energy ABS 
at the boundary,\cite{SF09} in agreement with the classification based on the invariant (\ref{Z2-invariant}). This state can also be viewed as two coupled Kitaev chains, see Ref. \onlinecite{Kit01}.
Since each chain contributes one zero-energy ABS, there are two zero modes at the end of the wire, corresponding to two Majorana quasiparticles.\cite{Majoranas} The stability of the 
Majorana states in the Kitaev chain and similar models against perturbations, such as interactions or disorder, has been investigated in a number of works, see Refs. \onlinecite{interactions-Z8} and \onlinecite{MF-stability}.

\section{Spectrum of the boundary modes}
\label{sec: ABS}

We consider a half-infinite superconductor at $x\geq 0$. The Bogoliubov quasiparticle wave function in each band is a two-component (electron-hole) spinor. 
In the vicinity of the Fermi point $rk_{F,\lambda}$, where $r=\pm$, it 
can be represented in the semiclassical approximation as $e^{irk_{F,\lambda}x}\psi_{\lambda,r}(x)$. The ``envelope'' function $\psi_{\lambda,r}$ varies slowly on the scale of the Fermi wavelength 
$k_{F,\lambda}^{-1}$ and satisfies the Andreev equation:\cite{And64}    
\begin{equation}
\label{And-eq-gen}
	\left(\begin{array}{cc}
		-iv_{\lambda,r}\dfrac{d}{dx} & \Delta_\lambda \\
		\Delta^*_\lambda & iv_{\lambda,r}\dfrac{d}{dx}
	\end{array}\right)\psi_{\lambda,r}=E\psi_{\lambda,r}.
\end{equation}
Here $v_{\lambda,r}=r\tilde v_F$ is the group velocity and $\Delta_\lambda\equiv\Delta_\lambda(k_{F,\lambda})=\Delta_\lambda(-k_{F,\lambda})$ 
is the gap function affecting the quasiparticles near the Fermi point $rk_{F,\lambda}$. To make analytical progress, we neglect self-consistency and assume that the gap functions do not depend on $x$.

We focus on the bound-state solutions of Eq. (\ref{And-eq-gen}) localized near $x=0$. The semiclassical approximation breaks down near the boundary due to a rapid variation of the potential, which leads to the mixing of 
the states corresponding to different Fermi wave vectors $\pm k_{F,\pm}$. As a result, the general wave function of the subgap states away from the boundary is given by a superposition 
of four semiclassical solutions:
\begin{equation}
\label{Psi-ABS}
  \Psi(x)=\sum_{\lambda=\pm}\sum_{r=\pm}\phi(rk_{F,\lambda})e^{-\Omega_\lambda x/\tilde v_F}e^{irk_{F,\lambda}x},
\end{equation}
where $\Omega_\lambda=\sqrt{|\Delta_\lambda|^2-E^2}$ and the Andreev amplitudes have the form
\begin{equation}
\label{Andreev amplitude}
	\phi(rk_{F,\lambda})=C(rk_{F,\lambda})\left(\begin{array}{c}
		\dfrac{\Delta_\lambda}{E-i\Omega_\lambda\sgn v_{\lambda,r}}\vspace*{5pt} \\ 1
	\end{array}\right).
\end{equation}
The ABS energy satisfies $|E|<\min(|\Delta_-|,|\Delta_+|)$.

Depending on the sign of the group velocity, the Fermi wave vectors in the wave function (\ref{Psi-ABS}) are classified as either incident, $k^{\mathrm{in}}_\lambda=-k_{F,\lambda}$, 
or reflected, $k^{\mathrm{out}}_\lambda=k_{F,\lambda}$. According to Eq. (\ref{Andreev amplitude}), the corresponding Andreev amplitudes are given by
\begin{equation}
\label{phi-in-out}
  \phi\left(k^{\mathrm{in(out)}}_\lambda\right)=C\left(k^{\mathrm{in(out)}}_\lambda\right)\left(\begin{array}{c}
		\dfrac{\Delta_\lambda}{E\pm i\Omega_\lambda}\vspace*{5pt} \\ 1
	\end{array}\right).
\end{equation}
The boundary condition for the wave function (\ref{Psi-ABS}) cannot be derived using the semiclassical approximation. 
As shown in Ref. \onlinecite{Shel-bc}, it can be written as a relation between the Andreev amplitudes for the reflected and incident states:
\begin{equation}
\label{Shelankov-bc}
  \phi(k^{\mathrm{out}}_\lambda)=\sum_{\lambda'} S_{\lambda\lambda'}\phi(k^{\mathrm{in}}_{\lambda'}).
\end{equation}
Here the coefficients $S_{\lambda\lambda'}$ form a unitary $2\times 2$ matrix ($S$ matrix), which is an electron-hole scalar, determined by the microscopic details of the boundary scattering 
at the Fermi level in the normal state. 
Inserting the expressions (\ref{phi-in-out}) into Eq. (\ref{Shelankov-bc}), we obtain the general form of the ABS energy equation, valid for any mechanism of the boundary scattering:
\begin{equation}
\label{ABS-equation}
  \frac{E^2-|\Delta_-||\Delta_+|\cos\chi}{\sqrt{(|\Delta_-|^2-E^2)(|\Delta_+|^2-E^2)}} = R,
\end{equation}
where
\begin{equation}
\label{R-def}
  R=1-2\frac{S_{--}S_{++}}{S_{-+}S_{+-}}=1+2\frac{|S_{--}|^2}{|S_{-+}|^2}
\end{equation}
(the second equality here follows from the unitarity of the $S$ matrix). As evident from Eq. (\ref{ABS-equation}), the ABS spectrum consists of symmetrical pairs $\pm|E|$; 
therefore the zero-energy states, if they exist, are twofold degenerate.

At $\chi=0$, the $\mathbb{Z}_2$ invariant (\ref{Z2-invariant}) is equal to $+1$, placing this bulk state into the topologically trivial class, without any stable zero modes. Indeed, in this case
the left-hand side of Eq. (\ref{ABS-equation}) is negative, while the right-hand side is positive, which means that there are no subgap ABS solutions, zero-energy or not, regardless of the boundary details. 
In the other TR invariant superconducting state, corresponding to $\chi=\pi$, we have $N_{1D}=-1$. However, it follows from Eqs. (\ref{ABS-equation}) and (\ref{R-def}) that the zero-energy
solution at $\chi=\pi$ exists only if $R=1$. This last condition is equivalent to the requirement that $S_{--}=S_{++}=0$, i.e., there is no backscattering into the same band. 
Thus we see that the zero modes are sensitive to the form of the boundary scattering matrix, even if the bulk state is topologically nontrivial.

\section{The $S$ matrix}
\label{sec: FM boundary}

In this section, we study the structure of the $S$ matrix in the presence of the TRS-breaking boundary scattering. Since the $S$ matrix is an electron-hole scalar, 
one can neglect the superconductivity and consider a normal metal in contact with a ferromagnetic insulator, the latter occupying the $x<0$ half-space. 
The ferromagnetism is modelled by the exchange splitting $h$ of the potentials affecting the two spin channels:    
\begin{equation}
\label{U-FM}
  \hat{U}_{FM}=U\hat\sigma_0+h\hat\sigma_3.
\end{equation}
We assume that $U\pm h>\epsilon_F$. The mismatch between the effective masses of quasiparticles on the metallic and insulating sides is neglected for analytical simplicity.

The normal-state Hamiltonian, naively written as $\hat{H}=\epsilon_0(\hat k_x)\hat{\sigma}_{0}+\theta(x)\bgam(\hat k_x)\hat{\bm{\sigma}}+\theta(-x)\hat{U}_{FM}$, where $\hat k_x=-i\nabla_x$, 
is not Hermitian due to the SO coupling term. To restore its Hermiticity, we use the following revised form:
$$
  \hat{H}=\epsilon_0(\hat k_x) \hat{\sigma}_{0} + \frac{1}{2}\{\bgam(\hat k_x),\theta(x)\}\hat{\bm{\sigma}} + \theta(-x)\hat{U}_{FM}, 
$$
where $\{...\}$ denotes the anticommutator and $\hat{U}_{FM}$ is given by Eq. (\ref{U-FM}). In the case of the 1D Rashba expression for the SO coupling, see Eq. (\ref{gamma-Rashba}), 
the Hamiltonian becomes
\begin{eqnarray}
\label{H-normal}
  \hat{H} &=& \epsilon_0(\hat k_x) \hat{\sigma}_{0} + \theta(-x)\hat{U}_{FM} + \theta(x)(\bm{a}\hat{\bm{\sigma}})\hat k_x\nonumber\\ 
          && +\frac{i}{2}(\bm{a}\hat{\bm{\sigma}})\delta(x), 
\end{eqnarray}
The additional delta-function term leads to a modification of the boundary conditions for the spinor wave functions:
\begin{equation}
\label{psi-bc}
  \left\{\begin{array}{l}
         \psi(+0)=\psi(-0)=\psi(0),\\
	  \psi'(+0)-\psi'(-0)=-i\Lambda\dfrac{(\bm{a}\hat{\bm{\sigma}})}{|\bm{a}|}\psi(0), 
         \end{array}\right.
\end{equation}
where $\Lambda=m^*|\bm{a}|$ characterizes the SO coupling strength.

The Fermi-level eigenstates of the Hamiltonian (\ref{H-normal}) are given by the evanescent waves on the insulating side ($x<0$): 
\begin{equation}
\label{FM-side}
  \psi(x)=C_\uparrow\left(\begin{array}{c}
                 1 \\ 0
                \end{array}\right)e^{\kappa_\uparrow x}
	  +C_\downarrow\left(\begin{array}{c}
                 0 \\ 1
                \end{array}\right)e^{\kappa_\downarrow x},
\end{equation}
with $\kappa_{\uparrow(\downarrow)}=\sqrt{2m^*(U\pm h-\epsilon_F)}$, and by a superposition of four propagating waves on the metallic side ($x>0$):
\begin{equation}
\label{metal-side}
  \psi(x)=\sum_{\lambda=\pm}\left(A_\lambda\chi_\lambda^{\mathrm{in}}e^{-ik_{F,\lambda} x}+B_\lambda\chi_\lambda^{\mathrm{out}}e^{ik_{F,\lambda} x}\right),
\end{equation}
with the Fermi wave vectors defined by Eq. (\ref{Fermi-wavevectors}). The spinor components of the propagating waves are obtained from Eq. (\ref{Rashba-eigenstates}): 
\begin{eqnarray*}
  \chi_-^{\mathrm{in}}=-i\chi_+^{\mathrm{out}}=-\frac{i}{\sqrt{\tilde v_F}}\left(\begin{array}{c}
                                                                       e^{-i\beta/2}\cos\dfrac{\alpha}{2}\vspace*{2pt} \\
                                                                       e^{i\beta/2}\sin\dfrac{\alpha}{2}
                                                                      \end{array}\right),\\
  \chi_-^{\mathrm{out}}=-i\chi_+^{\mathrm{in}}=-\frac{i}{\sqrt{\tilde v_F}}\left(\begin{array}{c}
                                                                       e^{-i\beta/2}\sin\dfrac{\alpha}{2}\vspace*{2pt} \\
                                                                       -e^{i\beta/2}\cos\dfrac{\alpha}{2}
                                                                  \end{array}\right),
\end{eqnarray*}
where we used the spherical angle parameterization of the SO coupling, see Eq. (\ref{a-angles}).

Substituting Eqs. (\ref{FM-side}) and (\ref{metal-side}) into the boundary conditions (\ref{psi-bc}), we can express the amplitudes of the reflected waves 
in terms of the amplitudes of the incident waves:
\begin{equation}
\label{S-matrix-def}
  B_\lambda=\sum_{\lambda'}S_{\lambda\lambda'}A_{\lambda'}.
\end{equation}
The scattering matrix here is given by
\begin{widetext}
\begin{equation}
\label{S-matrix-result}
  \hat S=-\frac{i}{(1+i\rho)^2+\rho_m^2}\left(\begin{array}{cc}
                2\rho_m\sin\alpha & -1-\rho^2+\rho_m^2-2i\rho_m\cos\alpha \vspace*{2pt} \\
               1+\rho^2-\rho_m^2-2i\rho_m\cos\alpha & 2\rho_m\sin\alpha
	  \end{array}\right),
\end{equation}
\end{widetext}
where 
$$
  \rho=\frac{\kappa_\uparrow+\kappa_\downarrow}{2\tilde k_F},\quad \rho_m=\frac{\kappa_\uparrow-\kappa_\downarrow}{2\tilde k_F}
$$
are the dimensionless measures of the TRS-preserving and TRS-breaking boundary scattering, respectively, satisfying the condition $|\rho_m|<\rho$.

If the boundary is nonmagnetic, i.e., $h=0$ and $\rho_m=0$, then Eq. (\ref{S-matrix-result}) takes the form
\begin{equation}
\label{S-nonmagnetic}
  \hat S=\frac{1-i\rho_0}{1+i\rho_0}\left(\begin{array}{cc}
                0 & i \\
               -i & 0
	  \end{array}\right),
\end{equation}
where $\rho_0=\sqrt{2m^*(U-\epsilon_F)}/\tilde k_F$.
In particular, in the case of an infinitely strong boundary potential, $U\to\infty$, we have $\hat S=\hat\sigma_2$ (note the difference from the dimensional 
reduction of the $S$ matrix for the 2D Rashba model,\cite{KS10} which is due to the different phase choice for the helicity eigenstates).  
The vanishing of the diagonal matrix elements in Eq. (\ref{S-nonmagnetic}) means that the backscattering into the same helicity band is forbidden by the TRS. 
This property does not in fact depend on the particular model of the boundary, as long as it is nonmagnetic, see Appendix \ref{app: S-offdiagonal}. 

If the boundary is magnetic, then there are no symmetry reasons for $S_{--}$ and $S_{++}$ to vanish, but it can happen by accident. 
For instance, regardless of the value of $\rho_m$, the diagonal elements in Eq. (\ref{S-matrix-result}) are equal to zero if $\sin\alpha=0$, 
which corresponds to $\bgam\parallel\hat{\bm{z}}$. However, one can show that such a configuration of the SO coupling is not protected by the point-group symmetry\cite{Sam17} 
and is unstable under a small perturbation of the confining potential.
Therefore, for the parameter $R$, which controls the boundary effects on the ABS spectrum, see Eq. (\ref{R-def}), we obtain that $R=1$ if and only if the boundary is nonmagnetic, otherwise $R>1$.
In particular, for the boundary model (\ref{H-normal}), the $S$ matrix is given by Eq. (\ref{S-matrix-result}) and we obtain:
\begin{equation}
\label{R-final}
R=1+\frac{8\rho_m^2\sin^2\alpha}{(1+\rho^2-\rho_m^2)^2+4\rho_m^2\cos^2\alpha}.
\end{equation} 
 
To summarize, the twofold degenerate ABS zero modes exist only if (i) the bulk superconducting state is real and topologically nontrivial, i.e., $\chi=\pi$, 
and (ii) the boundary is nonmagnetic, i.e., $S_{--}=S_{++}=0$ . Violation of either of these two conditions results in the gapping of the ABS.

\section{Stability of the zero modes}
\label{sec: ABS-stability}

\begin{figure}
\includegraphics[width=7cm]{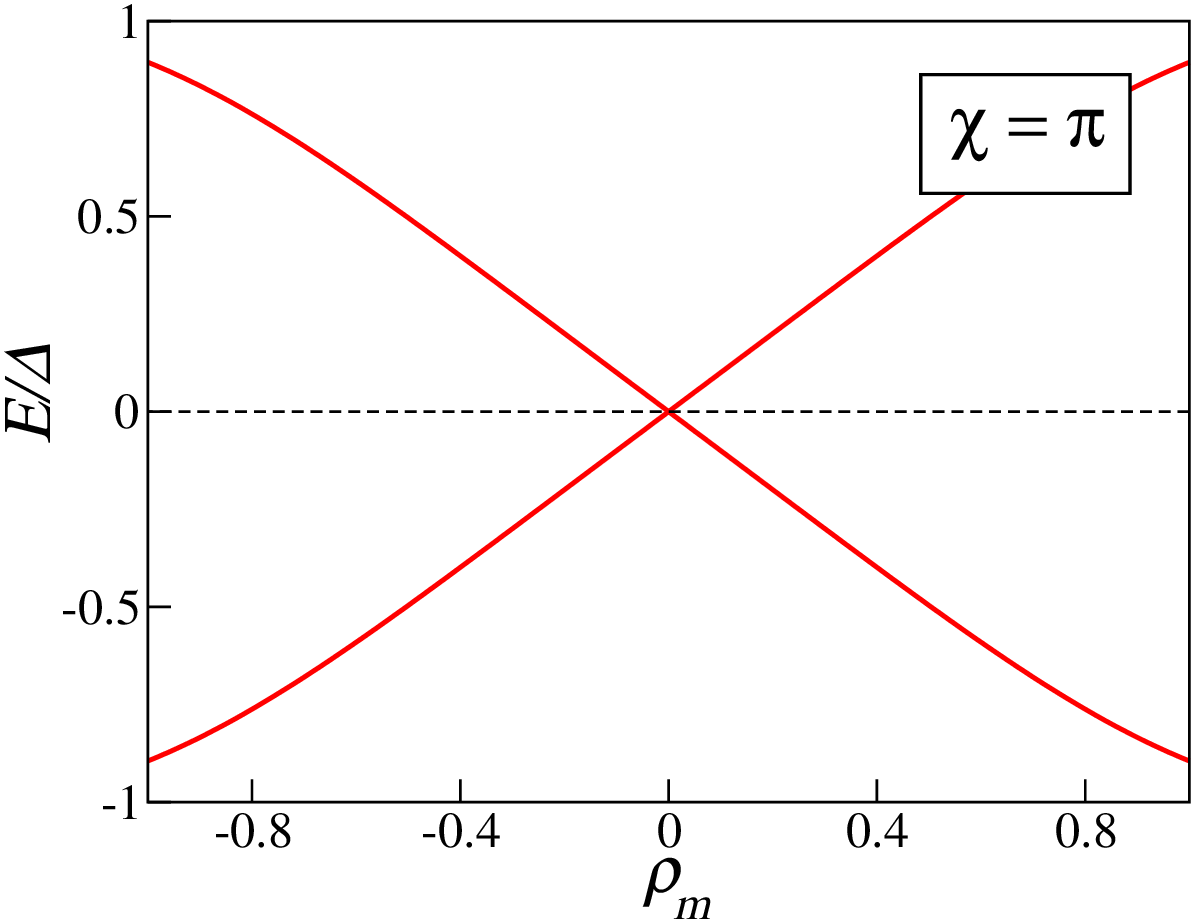}
\caption{The ABS energy as a function of the TRS-breaking boundary scattering, for $|\Delta_+|=|\Delta_-|=\Delta$, $\rho=1$, and $\alpha=\pi/2$.}
\label{fig: E-rho_m}
\end{figure}

\begin{figure}
\includegraphics[width=7cm]{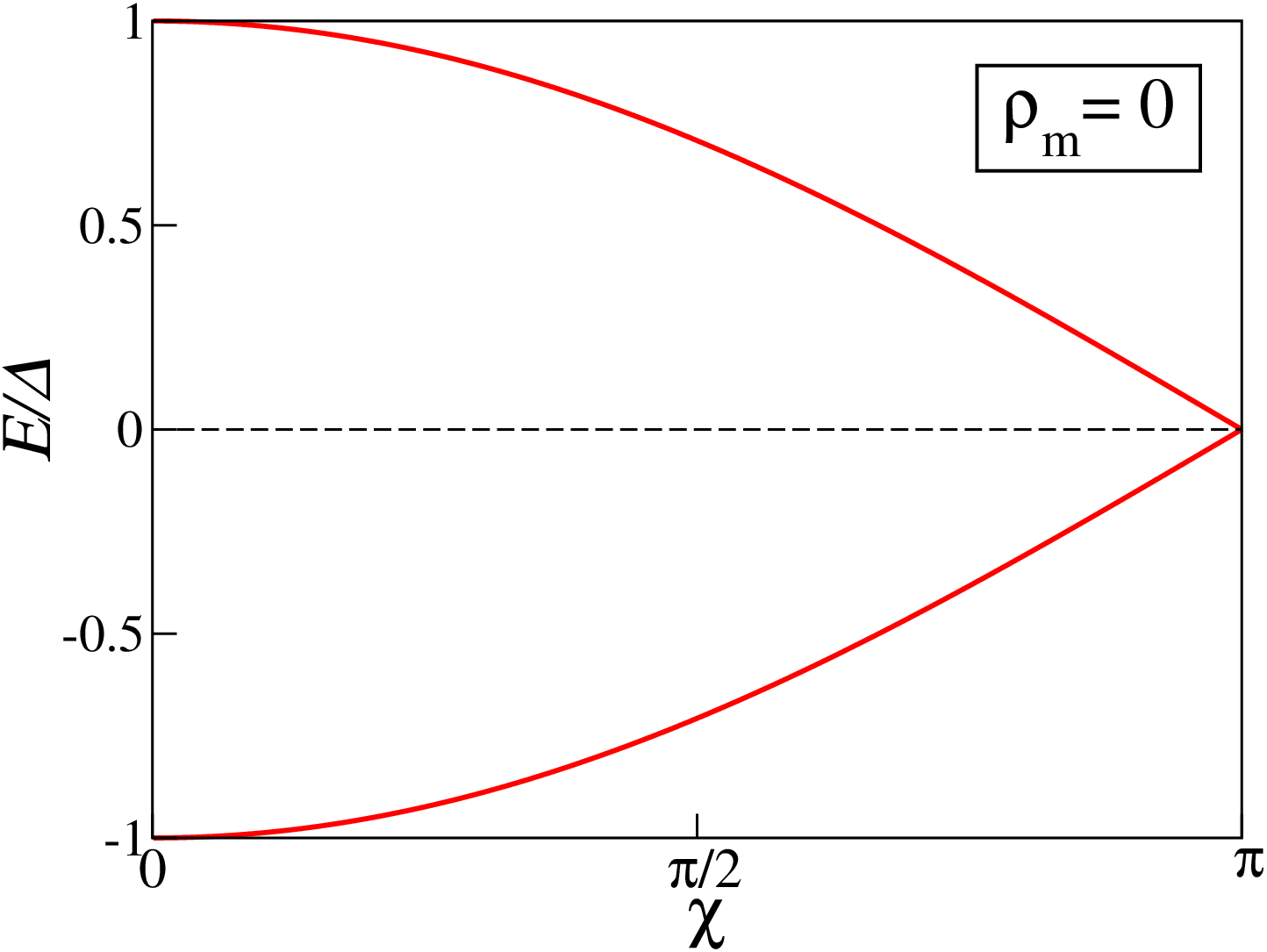}
\caption{The ABS energy as a function of the interband phase difference, for a nonmagnetic boundary and $|\Delta_+|=|\Delta_-|=\Delta$.}
\label{fig: E-chi}
\end{figure}

To illustrate the effects of the TRS-breaking perturbations on the ABS zero modes, we assume equal gap magnitudes in both helicity bands: $|\Delta_-|=|\Delta_+|=\Delta$, and obtain from Eq. (\ref{ABS-equation}) 
the following expression: $E=\pm\Delta\sqrt{(R+\cos\chi)/(R+1)}$,
where $R$ is given by Eq. (\ref{R-final}). If the bulk is TR invariant, with $\chi=\pi$, but the boundary scattering is not, then the ABS split and move to nonzero energies, as shown in Fig. \ref{fig: E-rho_m}. 
Any deviation of the interband phase difference from $\pi$ produces the same effect, see Fig. \ref{fig: E-chi}.

We see that the ABS are pushed apart and away from the zero energy if the mean-field state breaks TRS. It is then natural to ask if the zero modes are stable under TRS-breaking fluctuations.
We consider the superconducting state with arbitrary gap magnitudes, in which $h=0$ and $\chi=\pi$ at the mean-field level. However, both the exchange field (or magnetisation)
and the interband phase difference now experience classical thermal fluctuations:
$$
  h\to\delta h(x),\quad \chi\to\pi+\delta\chi(x).
$$ 
Here $\delta h$ and $\delta\chi$ are Gaussian-distributed random fields with zero average in the left and right half-spaces, respectively. 

For long-wavelength fluctuations, the ABS is affected only by the local values of the fluctuating fields at the boundary, $\delta h(0)$ and $\delta\chi(0)$. Then, it follows from Eqs. (\ref{R-final}) and (\ref{ABS-equation}) 
that $R=1+\delta R$ and
\begin{equation}
\label{E-squared-fluct}
  E^2=\frac{2|\Delta_-|^2|\Delta_+|^2}{(|\Delta_-|+|\Delta_+|)^2}\left[\delta R+\frac{1}{2}\delta\chi^2(0)\right],
\end{equation}
where
$$
  \delta R=2C^2\frac{\delta h^2(0)}{\epsilon^2_F},\quad C=\frac{\sin\alpha}{\rho_0(1+\rho_0^2)}\left(\frac{k_F}{\tilde k_F}\right)^2.
$$
Introducing the renormalized fluctuating fields 
$$
  m(x)=C\,\frac{\delta h(x)}{\epsilon_F},\qquad \varphi(x)=\frac{\delta\chi(x)}{2},
$$ 
and also the dimensionless energy 
$$
  \varepsilon=\frac{E}{\tilde\Delta},\qquad \tilde\Delta=\frac{2|\Delta_-||\Delta_+|}{|\Delta_-|+|\Delta_+|},
$$
we obtain from Eq. (\ref{E-squared-fluct}):
\begin{equation}
\label{conical}
  \varepsilon=\pm\sqrt{m^2(0)+\varphi^2(0)}.
\end{equation}
This form of the ABS energy dependence on the TRS-breaking perturbations is rather generic and can be understood as follows.   

The Bogoliubov quasiparticle spectrum is found by diagonalizing the BdG Hamiltonian ${\cal H}$, which incorporates the boundary potential and
can be written as ${\cal H}={\cal H}_{\mathrm{TRI}}+\delta{\cal H}$. The first term here is the TR invariant part, corresponding to $h=0$ and $\chi=0$ or $\pi$ in the model considered above. Its 
diagonalization in a topologically nontrivial state ($\chi=\pi$) produces two zero-energy ABS at each end of the superconducting wire. 
The $\delta{\cal H}$ term contains all TRS-breaking contributions, both from the bulk superconducting state and from the external fields, including a magnetic boundary. 
These contributions are assumed to be small and can therefore be treated by 
standard means of the degenerate pertubation theory. In the subspace of the zero modes localized near $x=0$,
the TRS-breaking Hamiltonian is represented by a Hermitian $2\times 2$ matrix $\bm{b}\hat{\bm{\tau}}$, where $\bm{b}$ is real and $\hat{\bm{\tau}}$ are the Pauli matrices, 
the unit matrix being absent due to the BdG electron-hole symmetry. The ABS energies are given by $\pm|\bm{b}|$, i.e. the zero-mode energy splitting has a nonanalytical, conical, dependence on the TRS-breaking perturbations,
in agreement with the explicit model calculation, see Eq. (\ref{conical}).

\subsection{ABS density of states}
\label{sec: ABS DOS}

The main quantity of interest for us is the fluctuation-averaged density of states (DoS) of the ABS, which is given  by the following expression:
$$
  N_{\mathrm{ABS}}(\varepsilon)=\left\langle\delta\left[\varepsilon-\sqrt{m^2(0)+\varphi^2(0)}\right]\right\rangle,
$$
at $\varepsilon\geq 0$, while $N_{\mathrm{ABS}}(-\varepsilon)=N_{\mathrm{ABS}}(\varepsilon)$. Introducing the notation $m(0)=m_0$ and $\varphi(0)=\varphi_0$, we obtain:
\begin{eqnarray}
\label{DOS-PP}
  N_{\mathrm{ABS}}(\varepsilon) = \iint_{-\infty}^{\infty}dm_0\,d\varphi_0\,P_m(m_0)P_\varphi(\varphi_0)\nonumber\\
   \times\delta(\varepsilon-\sqrt{m_0^2+\varphi_0^2}).
\end{eqnarray}
For analytical simplicity one can assume the Gaussian distributions of the local fluctuating fields: 
\begin{eqnarray}
\label{P-local}
  && P_m(m_0)=\frac{1}{\sqrt{2\pi\Gamma_m}}e^{-m_0^2/2\Gamma_m},\nonumber\\ \\
  && P_\varphi(\varphi_0)=\frac{1}{\sqrt{2\pi\Gamma_\varphi}}e^{-\varphi_0^2/2\Gamma_\varphi},\nonumber
\end{eqnarray}
where $\Gamma_m=\langle m^2(0)\rangle$ and $\Gamma_\varphi=\langle\varphi^2(0)\rangle$ characterize the magnitudes of the magnetic and the superconducting phase fluctuations, respectively.

In the absence of fluctuations, i.e., at $\Gamma_m=\Gamma_\varphi=0$, we have $P_m=\delta(m_0)$ and $P_\varphi=\delta(\varphi_0)$, and Eq. (\ref{DOS-PP}) yields 
\begin{equation}
\label{DOS-MF}
  N_{\mathrm{ABS}}(\varepsilon)=\delta(\varepsilon).
\end{equation}
If only one fluctuation channel is present, for instance, $\Gamma_m\neq 0$ but $\Gamma_\varphi=0$, then
\begin{equation}
\label{DOS-1-channel}
  N_{\mathrm{ABS}}(\varepsilon)=\frac{1}{\sqrt{2\pi\Gamma_m}}e^{-\varepsilon^2/2\Gamma_m},
\end{equation}
i.e., the delta-function peak in the DoS at zero energy is broadened.

\begin{figure}
\includegraphics[width=7cm]{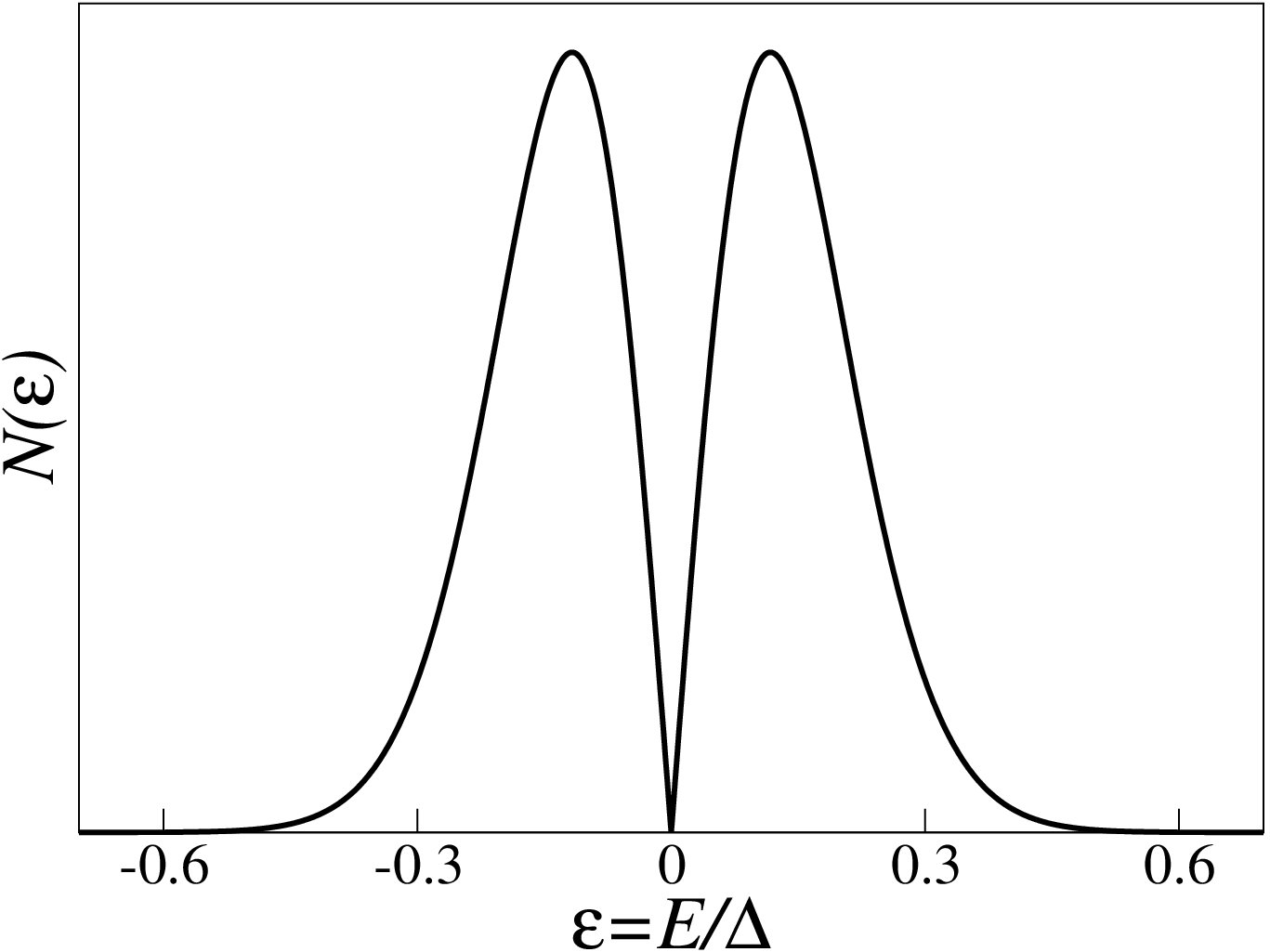}
\caption{The fluctuation-averaged ABS density of states, for $\Gamma_m=0.02$ and $\Gamma_\varphi=0.01$.}
\label{fig: DoS-2}
\end{figure}

In the general case, with both fluctuation channels taken into account, we obtain from Eq. (\ref{DOS-PP}):
\begin{eqnarray}
\label{DOS-2-channels}
  N_{\mathrm{ABS}}(\varepsilon) = \frac{\varepsilon}{\sqrt{\Gamma_m\Gamma_\varphi}}\exp\left[-\frac{\varepsilon^2}{4}\left(\frac{1}{\Gamma_m}+\frac{1}{\Gamma_\varphi}\right)\right]\nonumber\\
       \times I_0\left(\frac{\varepsilon^2}{4}\left|\frac{1}{\Gamma_m}-\frac{1}{\Gamma_\varphi}\right|\right),
\end{eqnarray}
where $I_0(z)$ is the modified Bessel function. Instead of the zero-energy peak, see Eqs. (\ref{DOS-MF}) and (\ref{DOS-1-channel}), the DoS now has a dip at low energies, with 
$N_{\mathrm{ABS}}(\varepsilon)\simeq\varepsilon/\sqrt{\Gamma_m\Gamma_\varphi}$ at $\varepsilon\to 0$, and two maxima symmetrically located at 
$\varepsilon\sim\sqrt{\Gamma_m\Gamma_\varphi/(\Gamma_m+\Gamma_\varphi)}$, see Fig. \ref{fig: DoS-2}. 
The most remarkable feature of the expresssion (\ref{DOS-2-channels}) is that 
\begin{equation}
\label{DoS-zero}
  N_{\mathrm{ABS}}(\varepsilon=0)=0,
\end{equation}
which means that the ABS zero modes are completely destroyed by the TRS-breaking fluctuations. 

It is easy to see that the result (\ref{DoS-zero}) holds true regardless of the details of the 
fluctuation distribution, as long as the latter is nonsingular. Indeed, at low energy one can replace the distribution functions in Eq. (\ref{DOS-PP}) by their values 
at $m_0=0$ and $\varphi_0=0$ and obtain:
$N_{\mathrm{ABS}}(\varepsilon\to 0)=P_m(0)P_\varphi(0)\iint dm_0\,d\varphi_0\,\delta(\varepsilon-\sqrt{m_0^2+\varphi_0^2})\propto\varepsilon$.

\section{Conclusions}
\label{sec: Conclusion}

We have calculated the spectrum of the Andreev boundary modes in a half-infinite superconducting wire on a substrate. This system is intrinsically noncentrosymmetric, which means that both the electron band structure and 
the Cooper pairing are strongly affected by the SO coupling of the Rashba type. We focused on the effects of two types of TR symmetry-breaking perturbations on the ABS spectrum.  While the external magnetic field is assumed to 
be zero, TRS can still be broken, either by a magnetic boundary scattering or intrinsically in the superconducting bulk, if the interband phase difference is not an integer multiple of $\pi$. We put the wire in contact with a 
ferromagnetic insulator and described the boundary scattering of the Andreev wave functions by the semiclassical $S$ matrix.   

We have shown that the symmetry and topology of the bulk Hamiltonian alone does not determine the number of the zero-energy ABS. 
Even if the $\mathbb{Z}_2$ topological invariant in the bulk points to the presence of a Kramers pair of the zero modes, the magnetic boundary scattering splits them. 
Thus the bulk-boundary correspondence, understood as a relation between the number of the zero modes and some topological characteristics of the bulk Hamiltonian, can break down if the symmetry of 
the boundary is lower than that of the bulk.

We have found that the zero modes are very sensitive to a ``virtual'' breaking of TRS. Even if the mean-field superconducting state is TR invariant and topologically nontrivial and the boundary is nonmagnetic, 
the TRS-breaking fluctuations lead to a complete vanishing of the fluctuation-averaged ABS DoS at zero energy. The overall shape of the DoS is qualitatively changed by the fluctuations, with a ``pseudogap'' 
minimum developing at low energies.

\acknowledgments
This work was supported by a Discovery Grant (K. S.) and by a USR Award (B. T.), both from the Natural Sciences and Engineering Research Council of Canada. 

\appendix

\section{TRS of the $S$ matrix}
\label{app: S-offdiagonal}

To prove that the helicity is reversed by a nonmagnetic boundary scattering, we write the wave function away from the boundary on the metallic side in the form  
\begin{equation}
\label{psi-general}
  |\psi\rangle=\sum_{\lambda=\pm}\left(A_\lambda|k^{\mathrm{in}}_\lambda,\lambda\rangle+B_\lambda|k^{\mathrm{out}}_\lambda,\lambda\rangle\right),
\end{equation}
cf. Eq. (\ref{metal-side}), where $|k,\lambda\rangle$ is the spinor Bloch state corresponding to the wave vector $k^{\mathrm{in}}_\lambda=-k_{F,\lambda}$ or $k^{\mathrm{out}}_\lambda=k_{F,\lambda}$. 
The $S$ matrix is defined by Eq. (\ref{S-matrix-def}). Applying the TR operation to the wave function (\ref{psi-general}) and using Eq. (\ref{t-factor}), we obtain: 
$$
 K|\psi\rangle=\sum_\lambda \left[B_\lambda^*t_\lambda(k^{\mathrm{out}}_\lambda)|k^{\mathrm{in}}_\lambda,\lambda\rangle
	+A_\lambda^*t_\lambda(k^{\mathrm{in}}_\lambda)|k^{\mathrm{out}}_\lambda,\lambda\rangle\right],
$$
with the ``in''- and ``out''-states interchanged by time reversal. Since the normal-state bulk Hamiltonian and the boundary are both nonmagnetic, one can expect the same 
$S$-matrix relations between the incident and reflected states in $|\psi\rangle$ and $K|\psi\rangle$, therefore
\begin{equation}
\label{TR-S-matrix}
  A_\lambda^*t_\lambda(k^{\mathrm{in}}_\lambda)=\sum_{\lambda'} S_{\lambda\lambda'}B_{\lambda'}^*t_{\lambda'}(k^{\mathrm{out}}_{\lambda'}).
\end{equation}
Comparing Eqs. (\ref{S-matrix-def}) and (\ref{TR-S-matrix}) and using the $S$-matrix unitarity, we arrive at the following constraint:
\begin{equation}
\label{S-TR-constraint}
  S_{\lambda'\lambda}=t_\lambda^*(k^{\mathrm{in}}_\lambda)S_{\lambda\lambda'}t_{\lambda'}(k^{\mathrm{out}}_{\lambda'}).
\end{equation}
Setting here $\lambda=\lambda'$ and using the property $t_\lambda(-k_x)=-t_\lambda(k_x)$, we obtain $S_{\lambda\lambda}=0$, regardless of the microscopic boundary details. 

Note that the $S$ matrix in 1D can be made antisymmetric by a suitable choice of the helicity eigenstates. Indeed, in the basis (\ref{Rashba-eigenstates}) the TR phase factors have the form (\ref{t-sgn-x}), 
therefore, $t_\lambda(k^{\mathrm{in}}_\lambda)=-1$ and $t_{\lambda}(k^{\mathrm{out}}_{\lambda})=+1$. Then, it follows from Eq. (\ref{S-TR-constraint}) that $S_{\lambda'\lambda'}=-S_{\lambda\lambda'}$.


\begin{thebibliography}{999}

\bibitem{top-SC}
B. A. Bernevig, {\it Topological Insulators and Topological Superconductors} (Princeton University Press, USA, 2013).

\bibitem{TKNN82}
D. J. Thouless, M. Kohmoto, M. P. Nightingale, and M. den Nijs, Phys. Rev. Lett. \textbf{49}, 405 (1982).

\bibitem{Volovik-book}
G. E. Volovik, \textit{The Universe in a Helium Droplet} (Clarendon Press, Oxford, 2003).

\bibitem{KM05}
C. L. Kane and E. J. Mele, Phys. Rev. Lett. \textbf{95}, 146802 (2005).

\bibitem{TI-review}
M. Z. Hasan and C. L. Kane, Rev. Mod. Phys. \textbf{82}, 3045 (2010).

\bibitem{SRFL08}
A. P. Schnyder, S. Ryu, A. Furusaki, and A. W. W. Ludwig, Phys. Rev. B \textbf{78}, 195125 (2008).

\bibitem{interactions-Z8}
L. Fidkowski and A. Kitaev, Phys. Rev. B \textbf{81}, 134509 (2010); E. Tang and X.-G. Wen, Phys. Rev. Lett. \textbf{109}, 096403 (2012).

\bibitem{QHE-edge}
B. I. Halperin, Phys. Rev. B \textbf{25}, 2185 (1982); Y. Hatsugai, Phys. Rev. Lett. \textbf{71}, 3697 (1993).

\bibitem{ABS-pwave}
T. L. Ho, J. R. Fulco, J. R. Schrieffer, and F. Wilczek, Phys. Rev. Lett. \textbf{52}, 1524 (1984); M. Matsumoto and M. Sigrist, J. Phys. Soc. Jpn. \textbf{68}, 994 (1999).

\bibitem{Hu94}
C.-R. Hu, Phys. Rev. Lett. \textbf{72}, 1526 (1994); S. Kashiwaya and Y. Tanaka, Rep. Prog. Phys. \textbf{63}, 1641 (2000).

\bibitem{BBC}
Y. Hatsugai, Phys. Rev. Lett. \textbf{71}, 3697 (1993); S. Ryu and Y. Hatsugai, Phys. Rev. Lett. \textbf{89}, 077002 (2002); V. Gurarie, Phys. Rev. B \textbf{83}, 085426 (2011);
G. M. Graf and M. Porta, Comm. Math. Phys. \textbf{324}, 851 (2013).

\bibitem{STYY11}
M. Sato, Y. Tanaka, K. Yada, and T. Yokoyama, Phys. Rev. B \textbf{83}, 224511 (2011).

\bibitem{MR15}
S. N. Molotkov and M. I. Ryzhkin, Pis'ma Zh. Eksp. Teor. Fiz. \textbf{102}, 216 (2015) [JETP Lett. \textbf{102}, 189 (2015)].

\bibitem{TRB-1D}
R. M. Lutchyn, J. D. Sau, and S. Das Sarma, Phys. Rev. Lett. \textbf{105}, 077001 (2010); Y. Oreg, G. Refael, and F. von Oppen, Phys. Rev. Lett. \textbf{105}, 177002 (2010).

\bibitem{Rashba-model}
Yu. A. Bychkov and E. I. Rashba,  Pis'ma Zh. Eksp. Teor. Fiz. \textbf{39}, 66 (1984) [JETP Lett. \textbf{39}, 78 (1984)].

\bibitem{Manchon15}
A. Manchon, H. C. Koo, J. Nitta, S. M. Frolov, and R. A. Duine, Nature Materials \textbf{14}, 871 (2015).

\bibitem{NCSC-book}
\textit{Non-centrosymmetric Superconductors: Introduction and Overview}, ed. by E. Bauer and M. Sigrist, Lecture Notes in Physics \textbf{847} (Springer, Heidelberg, 2012). 

\bibitem{Kneid15}
F. Kneidinger, E. Bauer, I. Zeiringer, P. Rogl, C. Blaas-Schenner, D. Reith, and R. Podloucky, Physica C \textbf{514}, 388 (2015).

\bibitem{Smid17}
M. Smidman, M. B. Salamon, H. Q. Yuan, and D. F. Agterberg, Rep. Prog. Phys. \textbf{80}, 036501 (2017). 

\bibitem{Sam17}
K. V. Samokhin, Phys. Rev. B \textbf{95}, 064504 (2017).

\bibitem{t-factor}
L. P. Gor'kov and E. I. Rashba, Phys. Rev. Lett. \textbf{87}, 037004 (2001); I. A. Sergienko and S. H. Curnoe, Phys. Rev. B \textbf{70}, 214510 (2004). 

\bibitem{TRSB-states}
D. F. Agterberg, V. Barzykin, and L. P. Gor'kov, Phys. Rev. B \textbf{60}, 14868 (1999); V. Stanev and Z. Te\v{s}anovi\'{c}, Phys. Rev. B \textbf{81}, 134522 (2010); 
Y. Tanaka, P. M. Shirage, and A. Iyo, Physica C \textbf{470}, 2023 (2010); 
S. Maiti and A. V. Chubukov, Phys. Rev. B \textbf{87}, 144511 (2013); V. Samokhin, Ann. Phys. (N.Y.) \textbf{359}, 385 (2015); K. V. Samokhin, Phys. Rev. B \textbf{92}, 174517 (2015).

\bibitem{TRI-1D}
C. L. M. Wong and K. T. Law, Phys. Rev. B \textbf{86}, 184516 (2012); F. Zhang, C. L. Kane, and E. J. Mele, Phys. Rev. Lett. \textbf{111}, 056402 (2013); 
E. Gaidamauskas, J. Paaske, and K. Flensberg, Phys. Rev. Lett. \textbf{112}, 126402 (2014); X.-J. Liu, C. L. M. Wong, and K. T. Law, Phys. Rev. X \textbf{4}, 021018 (2014); 
A. Haim, A. Keselman, E. Berg, and Y. Oreg, Phys. Rev. B \textbf{89}, 220504(R) (2014); A. Haim, E. Berg, K. Flensberg, and Y. Oreg, Phys. Rev. B \textbf{94}, 161110(R) (2016).

\bibitem{QHZ10}
X.-L. Qi, T. L. Hughes, and S.-C. Zhang, Phys. Rev. B \textbf{81}, 134508 (2010).

\bibitem{SF09}
M. Sato and S. Fujimoto, Phys. Rev. B \textbf{79}, 094504 (2009); Y. Tanaka, T. Yokoyama, A. V. Balatsky, and N. Nagaosa, Phys. Rev. B \textbf{79}, 060505(R) (2009).

\bibitem{Kit01}
A. Y. Kitaev, Physics-Uspekhi \textbf{44}, 131 (2001).  

\bibitem{Majoranas} 
J. Alicea, Rep. Prog. Phys. \textbf{75}, 076501 (2012); M. Leijnse and K. Flensberg, Semicond. Sci. Technol. \textbf{27}, 124003 (2012).

\bibitem{MF-stability}
M. Cheng and H.-H. Tu, Phys. Rev. B \textbf{84}, 094503 (2011); L. Fidkowski, R. M. Lutchyn, C. Nayak, and M. P. A. Fisher, Phys. Rev. B \textbf{84}, 195436 (2011); 
A. M. Lobos, R. M. Lutchyn, and S. Das Sarma, Phys. Rev. Lett. \textbf{109}, 146403 (2012); 
Y. Hu and M. A. Baranov, Phys. Rev. A \textbf{92}, 053615 (2015); N. M. Gergs, L. Fritz, and D. Schuricht, Phys. Rev. B \textbf{93}, 075129 (2016).

\bibitem{And64}
A. F. Andreev, Zh. Eksp. Teor. Fiz. \textbf{46}, 1823 (1964) [Sov. Phys.-JETP \textbf{19}, 1228 (1964)]; Ch. Bruder, Phys. Rev. B \textbf{41}, 4017 (1990); I. Adagideli, P. M. Goldbart, A. Shnirman, and A. Yazdani, 
Phys. Rev. Lett. \textbf{83}, 5571 (1999).

\bibitem{Shel-bc}
A. L. Shelankov, Pis'ma Zh. Eksp. Teor. Fiz. \textbf{32}, 122 (1980) [JETP Lett. \textbf{32}, 111 (1980)]; A. Millis, D. Rainer, and J. A. Sauls,
Phys. Rev. \textbf{B} 38, 4504 (1988); A. L. Shelankov and M. Ozana, Phys. Rev. B \textbf{61}, 7077 (2000).

\bibitem{KS10}
A. Khaetskii and E. Sukhorukov, JETP Letters \textbf{92}, 244 (2010).


\end{thebibliography}
\end{document}